\documentclass[lettersize,journal]{IEEEtran}
\IEEEoverridecommandlockouts
\usepackage{cite}
\usepackage{amsmath,amssymb,amsfonts}
\usepackage{graphicx}
\usepackage{textcomp}
\usepackage{xcolor}
\usepackage{booktabs}
\usepackage{array}
\usepackage{soul}
\usepackage{algorithm}
\usepackage{algpseudocode}
\usepackage{amsmath}
\usepackage{multirow}
\usepackage[caption=false,font=footnotesize]{subfig}  
\usepackage{graphicx}
\usepackage{url}

\def\BibTeX{{\rm B\kern-.05em{\sc i\kern-.025em b}\kern-.08em
    T\kern-.1667em\lower.7ex\hbox{E}\kern-.125emX}}
\begin{document}

\title{Adaptive Semantic Communication with Residual Quantization for Resilient Vehicular Networks}

\author{
		\IEEEauthorblockN{Xuanhao Luo,
            Ruichen Gao,
            Sandeep Sreenivasan, 
            Zhizhen Li,
            Yuchen Liu,~\IEEEmembership{Member,~IEEE}
 }

\thanks{X. Luo, S. Sandeep, Z. Li and Y. Liu are with the Department of Computer Science, North Carolina State University, Raleigh, NC, 27695, USA (Emails: \{xluo26, srsreeni, zli92, yuchen.liu\}@ncsu.edu).}
\thanks{R. Gao is with the Department of Electrical and Computer Engineering, New York University, Brooklyn, NY, 11201, USA {(Email: ruichen.gao@nyu.edu).}}
}

\markboth{Journal of \LaTeX\ Class Files,~Vol.~14, No.~8, 2025}%
{Shell \MakeLowercase{\textit{et al.}}: A Sample Article Using IEEEtran.cls for IEEE Journals}

\maketitle

\begin{abstract}
Vehicular networks require timely visual information exchange to support safety-critical applications such as cooperative perception and hazard awareness, yet transmitting high-dimensional camera data over bandwidth-limited and dynamic wireless links remains challenging. 
{In this paper, we propose an adaptive semantic communication architecture for vehicular multicast that adjusts the amount of visual information transmitted according to current channel conditions, available communication resources, and latency requirements. Specifically, the visual encoder uses residual quantization to represent each image at multiple bitrate and reconstruction-quality levels. At runtime, the system selects an appropriate level based on the communication conditions, enabling it to trade visual fidelity for lower transmission overhead when network resources are limited. Context-adaptive entropy coding further compresses the quantized representations using offline-learned statistics, without requiring retraining for different channel conditions. This design allows the visual representation to be learned offline while the communication strategy is adapted online. Experimental results demonstrate competitive rate-distortion performance at ultra-low bitrates, improved delivery reliability and deadline satisfaction across diverse vehicular scenarios, and strong downstream perception performance, achieving up to 0.88 mAP50 while dynamically adapting to changing communication conditions.}
\end{abstract}


\section{Introduction}

\label{sec:intro}
Modern visual communication systems, such as vehicular perception \cite{ashraf2021see}, satellite imaging \cite{li2025adaorb}, and edge–cloud vision platforms \cite{zhang2019hetero}, require the efficient transmission of high-dimensional visual data over bandwidth-limited and dynamically varying wireless links. In these scenarios, camera or LiDAR data must be delivered in real time despite rapid fluctuations in channel conditions caused by mobility, interference, and environmental dynamics. Vehicular networks, in particular, represent a highly demanding setting, where cooperative perception \cite{bista2018semantic}, collision avoidance \cite{luo2023clothoid}, and emergency response \cite{zhong2023secure} rely on low-latency and reliable dissemination of visual information \cite{gao2022communication}. However, road wireless environments are inherently dynamic, with rapidly varying channel conditions, heterogeneous receiver populations, and limited communication resources. These factors make it challenging to deliver high-dimensional visual data efficiently.
Conventionally, visual communication pipelines rely on pixel-level codecs such as JPEG or HEVC to reduce transmission overhead. While effective in removing statistical redundancy, these methods are agnostic to task semantics, allocating bits uniformly across the scene regardless of their relevance to downstream perception tasks.

This limitation motivates the development of semantic communication (SemCom) techniques, which focus on preserving task relevant information to achieve higher compression efficiency \cite{getu2025semantic}. Many recent works adopt deep joint source–channel coding (JSCC) \cite{xu2023deep}, which jointly optimizes source representation and channel transmission under specific physical layer settings.
However, such designs inherently couple source and channel coding, making their performance dependent on the channel assumptions considered during training. When channel conditions deviate from the training regime, adaptation or retraining may be required \cite{tang2026cache}, and reconstruction quality can degrade as the channel quality deteriorates \cite{bourtsoulatze2019deep}. This coupling is less suited to our target setting of safety critical vehicular communication, where predictable reconstruction after successful delivery and flexible runtime adaptation under rapidly varying channel and resource conditions are particularly important.
Meanwhile, some generative model based approaches \cite{tang2024evolving, tang2025enabling} reconstruct data by sampling from learned data priors, which may produce perceptually plausible but semantically inconsistent outputs. Such behavior can undermine the reliability of reconstructed content in safety critical vehicular perception tasks, for instance, when nearby vehicles are missed, blurred, or spatially distorted.
More importantly, existing works primarily focus on reconstruction or task performance under channel impairments, while transmission latency, delivery reliability, downstream perception utility, and runtime adaptation must be jointly considered in dynamic vehicular networks.

Motivated by these challenges, we adopt a system-oriented perspective and decouple semantic source coding from channel modeling to improve communication \textit{flexibility} and \textit{adaptability}. 
Here, flexibility refers to reuse pretrained semantic representations across heterogeneous and evolving wireless environments without requiring redesign or retraining for specific channel models. Adaptability, on the other hand, captures the capability of the system to dynamically adjust transmission strategies at runtime, such as rate allocation, redundancy control, or semantic prioritization, in response to time-varying channel conditions and task requirements, thereby maintaining consistent end-task performance under uncertainty. 
Our design therefore focuses on learning compact and structured visual representations that enable predictable and controllable trade-offs among semantic fidelity, transmission latency, and delivery reliability.

{To this end, we develop an adaptive SemCom architecture, which integrates residual quantization (RQ) with n-gram-based arithmetic coding (NAC). The RQ module exposes multiple hierarchical discrete representations with different quantization depths, allowing the residual quantization depth to serve as a runtime communication control variable across different bitrate-fidelity operating points. The NAC module then losslessly compresses the discrete index sequence at each operating point by exploiting short-range contextual dependencies, further reducing the transmission payload without altering the underlying semantic representation.}
To address time-varying wireless conditions, we then develop a lightweight model adaptation mechanism that dynamically adjusts the semantic compression depth to meet varying bandwidth and latency constraints. It is worth noting that the proposed framework is channel-agnostic at the representation learning stage, while remaining channel-aware at runtime through lightweight depth adaptation in the second stage. 
Beyond conventional reconstruction metrics, we jointly consider perceptual task accuracy, delivery success, and latency. This system-level evaluation framework provides a comprehensive assessment of whether transmitted visual information remains useful under realistic communication and computing constraints. By combining adaptive semantic compression with perception-aware evaluation, RQ-NAC bridges the gap between communication efficiency and semantic reliability in safety-critical vehicular networks.
The main contributions of this paper are summarized as follows:
\begin{itemize}

    \item {We develop a system-level adaptive semantic communication architecture for vehicular multicast that exposes pretrained RQ configurations as runtime-selectable operating points, enabling the communication system to adapt the transmitted representation to dynamic channel, resource, and latency conditions without online retraining.}

    \item {We propose a scalable compression framework, RQ-NAC, that combines residual quantization with context-adaptive lossless entropy coding. Different RQ depth configurations provide controllable bitrate and reconstruction fidelity trade-offs, while the offline entropy model further reduces the corresponding residual index streams.}

    \item {We design a lightweight runtime depth-selection mechanism that jointly considers effective multicast capacity, receiver scale, and application deadlines. The controller dynamically selects the RQ operating point under a unified policy, enabling adaptive trade-offs between reconstruction fidelity and timely delivery.}

    \item {We conduct a comprehensive system-level evaluation across representative vehicular multicast scenarios, including rate-distortion comparison with recent learning-based compression baselines, runtime profiling, controlled channel and resource sensitivity, downstream perception, and out-of-distribution robustness.}
\end{itemize}

{The remainder of this paper is organized as follows. Section~II reviews related work. Section~III introduces the vehicular communication model and operating scenarios. Section~IV presents the proposed RQ-NAC framework, and Section~V describes the runtime adaptive depth-selection mechanism. Section~VI presents the evaluation methodology, while Section~VII reports the experimental results on compression, system-level adaptation, downstream perception, and robustness under distribution shift. Section~VIII concludes the paper.}

\begin{figure*}[t]
    \centerline{\includegraphics[scale=1.04]{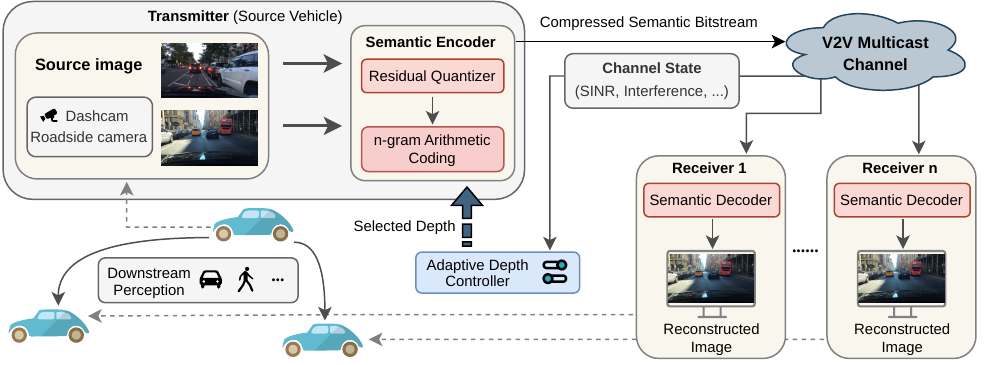}}
    \caption{System overview of the proposed adaptive SemCom framework for V2V multicast in wireless networks. A source vehicle encodes visual data from its camera or sensor into a compact semantic representation, which is then transmitted to multiple receivers over a shared wireless channel. The compression configuration is dynamically adjusted based on channel conditions and system constraints.}
    \label{v2v}
\end{figure*}

\section{Related Work}
Network planning and communication under limited resources remain fundamental challenges in dynamic wireless systems \cite{guo2019resource, wang2020learn, luo2026sparsepilot}.
Semantic communication has emerged as a promising paradigm to address the limitations of traditional bit-level communication, particularly in resource-constrained and dynamic environments such as vehicular networks \cite{xu2023semantic, du2025task, yuan2025learning}. The core idea is to transmit only the task-relevant information, thereby improving communication efficiency and robustness \cite{yang2022semantic, guo2024survey}. 
In general, SemCom systems can be broadly categorized into task-oriented and reconstruction-oriented paradigms based on their transmission objectives.
Task-oriented SemCom focuses on transmitting task-relevant features for downstream applications, such as image classification, machine translation, and visual question answering~\cite{xie2022task}.
In contrast, reconstruction-oriented SemCom aims to reconstruct semantically faithful contents or signals at the receiver \cite{lokumarambage2023wireless}. In a broader sense, reconstruction can also be regarded as a specific type of task \cite{qin2021semantic} in communication systems.

Within these SemCom paradigms, JSCC has been widely studied as a representative approach for end-to-end transmission \cite{xu2023deep, bourtsoulatze2019deep, gunduz2024joint, yang2021deep}. Classical JSCC schemes are typically tailored to specific source–channel pairs, limiting their generalizability across different communication environments \cite{gunduz2024joint}. Recent deep learning-based JSCC approaches leverage neural networks to learn end-to-end mappings, demonstrating improved robustness under noisy conditions \cite{xu2023deep, bourtsoulatze2019deep}. {HARQ-assisted SemCom has also been explored for LiDAR cooperative perception under noisy wireless channels~\cite{sheng2025semantic}. 
These approaches primarily focus on preserving semantic or task information under channel impairments. In contrast, our work targets scalable ultra low bitrate visual source coding for bandwidth constrained vehicular communication, while decoupling representation learning from the channel model and adapting the RQ depth at runtime according to communication and latency constraints. This distinction is particularly relevant to safety critical vehicular applications, where reliability and timely delivery are essential system requirements~\cite{pan2025trustworthy}. In such settings, the decoupled design avoids direct channel induced degradation of the reconstructed representation after successful delivery.}

To reduce reliance on end-to-end training and improve adaptability in dynamic wireless environments, recent works explore generative model-based SemCom. For example, \cite{tang2024evolving, tang2026cache} leverage generative adversarial networks (GAN) inversion with pretrained generative models to enable training-free semantic transmission. \cite{tang2025enabling} proposes a training-free generative SemCom system using diffusion models, where semantic features are transmitted and images are reconstructed at the receiver via noise-aware diffusion sampling. While such approaches improve perceptual quality and flexibility, generative models such as GANs, variational autoencoder (VAEs), and diffusion models reconstruct data by sampling from learned data priors \cite{goodfellow2014generative, kingma2013auto, ho2020denoising}, which may produce perceptually plausible but semantically inconsistent outputs, i.e., hallucinations. This prior-driven generation leads to semantic drift from the original input, which is undesirable in safety-critical applications such as vehicular perception, where reliable and accurate perception is essential \cite{pan2024reliability}. This limitation motivates the exploration of more reliable semantic representation schemes.

To address the above limitations of generative model-based approaches, recent works have explored alternative designs based on discrete latent representations that \textbf{decouple} source and channel coding. In \cite{hu2023robust}, a framework based on vector-quantized VAE (VQ-VAE) is adopted to encode images into discrete latent codes, improving reliability and compatibility with digital communication systems. Unlike generative approaches, discrete representation-based methods perform deterministic encoding and reconstruction, preserving structural consistency with the original input while introducing only controlled distortion. From a probabilistic perspective, VQ-VAE can still be interpreted within the VAE framework, where the posterior becomes deterministic and the Kullback–Leibler (KL) divergence term reduces to a constant under a uniform prior \cite{van2017neural}.
Recent efforts further improve VQ-based compression by enhancing the quantization process. For instance, Adaptive VQ-VAE \cite{chen2023adaptive} introduces a hyper-network to generate adaptive side information, which reduces quantization distortion and improves rate-distortion performance. However, such approaches still rely on single-stage vector quantization, limiting their representational capacity and scalability, especially under highly constrained bitrate regimes.

Residual-quantized VAE (RQ-VAE) \cite{lee2022autoregressive} further extends VQ-VAE via hierarchical residual quantization, enabling multi-stage refinement of latent representations with controllable fidelity and bitrate. This hierarchical structure provides greater flexibility and expressiveness compared to conventional VQ-based methods, making it particularly suitable for scalable semantic compression under dynamic wireless conditions.
In our preliminary work \cite{luo2026channel}, we introduced RQ-NAC as a channel-agnostic semantic compression framework that combines residual quantization with n-gram-based arithmetic coding for bandwidth-limited visual communication. This paper substantially extends that framework by introducing runtime depth adaptation under multicast channel and latency constraints, block-level HARQ modeling, system-level reliability and latency evaluation,  controlled channel and resource sensitivity studies, and downstream perception analysis. These extensions move the focus from standalone semantic compression to an adaptive SemCom system for dynamic vehicular networks.

\section{{System Model}}
\label{sec:system_model}

Fig. 1 illustrates the overall vehicle-to-vehicle (V2V) semantic communication framework. At the source vehicle, visual data are encoded into a compact semantic bitstream and transmitted to multiple receivers over a shared wireless channel. Each receiver decodes the received bitstream to reconstruct the visual content for downstream perception. An adaptive depth controller at the transmitter selects the RQ operating point according to the observed channel state and system constraints, enabling trade-offs among compression efficiency, transmission latency, and reconstruction quality.

\subsection{Communication Model}

Let $\mathcal{V} = \{v_1, v_2, \dots, v_N\}$ denote the set of receivers within a local communication region, and let $\{\gamma_u\}_{u=1}^N$ denote their instantaneous signal-to-interference-plus-noise ratio (SINR) values. To model reliable multicast under heterogeneous receiver conditions, we adopt a coverage-based bottleneck abstraction and define the effective SINR as the $(1-\eta)$-quantile of the receiver SINR distribution:
\begin{equation}
\gamma_\eta = \mathrm{Quantile}_{1-\eta}(\{\gamma_u\}),
\end{equation}
where $\eta$ denotes the target receiver coverage ratio. The effective SINR is then mapped to the multicast rate $R_{\mathrm{mc}}$ through the PHY abstraction described in Section~VI-D.

With $K$ concurrent multicast flows and a resource fraction $\rho \in (0,1]$ allocated to visual semantic dissemination, the effective resource fraction available to each flow is $\alpha_{\mathrm{eff}} = \frac{\rho}{K}$.
Assuming an orthogonal frequency-division multiplexing (OFDM)-based physical layer with $N_{\mathrm{sc}}^{\mathrm{full}}$ available subcarriers per scheduling interval, the effective number of subcarriers assigned to each multicast flow is
\begin{equation}
N_{\mathrm{sc}} = \alpha_{\mathrm{eff}} N_{\mathrm{sc}}^{\mathrm{full}},
\end{equation}
which determines the achievable multicast throughput under the PHY abstraction. 
Accordingly, the effective multicast rate $R_{\mathrm{mc}}$ is jointly determined by the effective SINR $\gamma_\eta$ and the allocated resource fraction $\alpha_{\mathrm{eff}}$.

\subsection{Block Transmission and HARQ Model}

To accurately capture the end-to-end latency and reliability of semantic information delivery, we model the transmission process using a block-based abstraction inspired by transport-block-level HARQ in 5G NR systems \cite{3gpp2022nr}. The compressed bitstream of size $S$ bits is segmented into fixed-size transmission blocks of $S_{\mathrm{blk}}$ bits, each corresponding to a PHY-layer transport block that can be independently decoded and retransmitted.

Under the PHY abstraction, the number of successfully decoded bits per scheduling interval is denoted by $b_{\mathrm{slot}}$, which is determined by the effective SINR $\gamma_\eta$ and resource allocation $\alpha_{\mathrm{eff}}$. The transmission time of a single block is given by
\begin{equation}
T_{\mathrm{blk}} =
\left\lceil \frac{S_{\mathrm{blk}}}{b_{\mathrm{slot}}} \right\rceil T_{\mathrm{slot}}.
\end{equation}

We adopt a simplified HARQ model in which each transmission block can be retransmitted up to $N_{\mathrm{retx}}$ times upon decoding failure. Let $n_i$ denote the number of transmission attempts for the $i$-th block. The total transmission latency is then expressed as
\begin{equation}
T_{\mathrm{tx}} = \sum_{i=1}^{M} n_i \, T_{\mathrm{blk}},
\end{equation}
where $M = \lceil S / S_{\mathrm{blk}} \rceil$ is the total number of blocks.
The resulting transmission latency is incorporated into the end-to-end latency model used for runtime depth selection in Section V.

\subsection{Representative Operating Scenarios}
To capture heterogeneous traffic and communication requirements in vehicular networks, we consider three representative operating scenarios: Cruise, Junction, and Collision. These scenarios differ in multicast scale, channel conditions, communication load, and latency requirements, providing a range of operating conditions from relatively relaxed to highly constrained transmission. The detailed parameter settings are given in Section VI-D.

\begin{figure*}[t]
	\centerline{\includegraphics[scale=1.2]{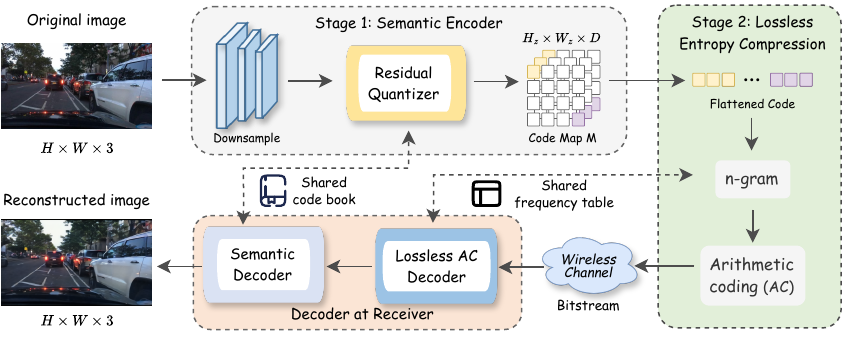}}
    \caption{Overview of the proposed RQ-NAC pipeline, consisting of a residual-quantized semantic encoder and a context-adaptive entropy coding module.} 
    \label{framework}
\end{figure*}

\section{Proposed Two-Stage Semantic Compression Framework}

The core component of the proposed semantic communication system is the semantic encoder shown in Fig.~1. As illustrated in Fig.~\ref{framework}, the proposed solution, called RQ-NAC, consists of a residual-quantization-based semantic encoder followed by a lossless context-adaptive entropy coding module, forming a two-stage pipeline for scalable semantic compression and efficient transmission.

\subsection{Residual Quantization for Semantic Compression}

In the first stage, a semantic encoder transforms the input visual data into hierarchical discrete latent representations with controllable quantization depth. 
Formally, let $E(\cdot)$ and $G(\cdot)$ denote the encoder and decoder, respectively. 
Given an input image $X \in \mathbb{R}^{H \times W \times 3}$, the encoder produces a latent feature map:
\begin{equation}
Z = E(X) \in \mathbb{R}^{H_z \times W_z \times n_z}, 
\end{equation}
where $H_z$ and $W_z$ are the spatial dimensions of the latent space, and $n_z$ is the feature dimensionality per spatial location. 
Each spatial feature vector $Z_{h,w} \in \mathbb{R}^{n_z}$ is quantized using residual quantization with depth $D$ and a shared codebook 
\begin{equation}
\mathcal{C} = \{ e(1), e(2), \ldots, e(K) \}, 
\end{equation}
where $e(k) \in \mathbb{R}^{n_z}$ denotes the $k$-th codeword, and $K$ is the codebook size.

At the stage $d$, the quantizer selects a codeword $e(k^{(d)}_{h,w})$ from the
codebook and refines the quantized approximation:
\begin{equation}
    \hat{Z}^{(d)}_{h,w} = \sum_{i=1}^{d} e(k^{(i)}_{h,w}), 
    \quad
    \hat{Z}_{h,w} = \hat{Z}^{(D)}_{h,w},
\end{equation}
where $d \in \{1,\ldots,D\}$ denotes the residual quantization depth.
The selected codeword indices are
\[
    M_{h,w} = \big(k^{(1)}_{h,w}, \ldots, k^{(D)}_{h,w}\big),
\]
which constitutes the discrete symbols generated during quantization. Collectively, 
these index sequences form the code map 
$M \in [K]^{H_z \times W_z \times D}$, which serves as the compact discrete 
representation transmitted over the wireless channel.
At the receiver side, the transmitted code map $M$ is used to reconstruct the 
quantized latent map $\hat{Z} \in \mathbb{R}^{H_z \times W_z \times n_z}$, 
which is subsequently decoded to obtain the final image $ \hat{X} = G(\hat{Z})$.

{The overall training objective combines a reconstruction term and a multi-level commitment term as:
\begin{equation}
    \mathcal{L} 
    = \|X - \hat{X}\|_2^{2}
    + \beta \sum_{d=1}^{D} \|Z - \text{sg}[\hat{Z}^{(d)}]\|_2^{2},
\end{equation}
where $\text{sg}[\cdot]$ denotes the stop-gradient operator, which acts as the identity function during the forward pass while blocking gradient propagation during backpropagation~\cite{van2017neural}.}
Following the RQ-VAE implementation, we update the codebook using exponential moving average (EMA) of the clustered encoder outputs instead of gradient-based updates. This approach stabilizes training and prevents codebook collapse. Additionally, perceptual loss and patch-level adversarial loss are incorporated to enhance visual fidelity.  This hierarchical quantization structure refines the latent representation level by level while sharing the same compact codebook across all depths, producing a discrete index tensor $M$.
We then flatten the index tensor $M$ into a one-dimensional sequence $\mathbf{s} = (s_1, s_2, \dots, s_N)$ with length $N = H_z W_z D$. This sequence serves as the input to the entropy coding stage as follows.

\subsection{Context-adaptive Arithmetic Coding}

After the semantic compression stage, the residual quantization encoder outputs a sequence of discrete indices 
$\mathbf{s}$. 
Although already compact, this sequence still contains statistical redundancy due to the strong correlation among neighboring indices. To further exploit these dependencies, we adopt a context-adaptive lossless entropy coding scheme.

From an information-theoretic view, the minimum achievable code length of this sequence is its conditional entropy:
\begin{equation}
    H(S \mid C) = -\sum_{s_t,\mathbf{c}_t} p(s_t,\mathbf{c}_t)\,\log p(s_t \mid \mathbf{c}_t),
\end{equation}
where $p(s_t,\mathbf{c}_t)$ is the joint distribution of the symbol and its context.
In practice, arithmetic coding uses an estimated model $q(\cdot)$, leading to an expected coding rate given by the cross-entropy $ H(p,q) = \mathbb{E}_{p(s_t,\mathbf{c}_t)}\!\big[-\log q(s_t \mid \mathbf{c}_t)\big].$
Thus, improving the accuracy of $q$ directly tightens the gap between the achievable rate and the entropy bound.

To better approximate $p(\cdot)$ and capture local dependencies, we employ an $n$-gram–based probabilistic model.  
Although predictive autoregressive models such as LLMs or transformer-based entropy models are used in natural language \cite{valmeekam2023llmzip} and byte-level \cite{luo2025rank, luo2025unified} compression, their sequential inference is too slow for real-time visual data transmission.
Thus, the $n$-gram model provides a low computational overhead while still capturing essential dependencies for entropy coding.
Given a context $\mathbf{c}_t = (s_{t-n+1}, \dots, s_{t-1})$, the conditional probability of the next symbol is estimated by maximum likelihood:
\begin{equation}
    \hat{P}(s_t \mid \mathbf{c}_t) = 
    \frac{N(\mathbf{c}_t, s_t)}{N(\mathbf{c}_t)},
\end{equation}
where $N(\mathbf{c}_t, s_t)$ denotes the number of occurrences of the $(\mathbf{c}_t, s_t)$ pattern in the training corpus, and $N(\mathbf{c}_t)$ is the number of occurrences of the context $\mathbf{c}_t$.  
By conditioning on $\mathbf{c}_t$, the $n$-gram estimator provides a more faithful approximation of the true sequence statistics, thereby reducing the cross-entropy term $H(p,q)$ and improving lossless compression efficiency.

However, when certain $n$-gram combinations are unseen during training, direct estimation may yield zero probabilities, which can collapse the arithmetic coding (AC) interval. 
To mitigate this issue, we apply additive smoothing \cite{chen1999empirical} that assigns a small nonzero probability to unseen symbols. 
The smoothed conditional probability is defined as:
\begin{equation}
    \tilde{P}(s_t \mid \mathbf{c}_t) = 
    \frac{N(\mathbf{c}_t, s_t) + \alpha}{N(\mathbf{c}_t) + \alpha K},
\end{equation}
where $\alpha$ is a small positive smoothing coefficient ($0 < \alpha \leq 1$) and $K$ is the codebook size.  
This approach ensures that every possible symbol maintains a nonzero probability, thereby preventing numerical instability and guaranteeing robust encoding and decoding across all contexts.
In our framework, the $n$-gram estimator $\tilde{P}(s_t \mid \mathbf{c}_t)$ serves as the conditional probability model $q(s_t \mid \mathbf{c}_t)$ used by AC.
The $n$-gram frequency table is constructed offline and shared between encoder and decoder, so no additional side information needs to be transmitted.

After obtaining the probability distribution from the n-gram model, AC progressively narrows an interval $[l,h)$ to encode the entire sequence.
At step $t$, the interval is updated according to the cumulative distribution function $F(\cdot \mid \mathbf{c}_t)$ of $\tilde{P}(\cdot \mid \mathbf{c}_t)$:
\begin{equation}
\begin{aligned}
    l_t &= l_{t-1} + (h_{t-1} - l_{t-1}) \cdot F(s_t \mid \mathbf{c}_t), \\
    h_t &= l_{t-1} + (h_{t-1} - l_{t-1}) \cdot F(s_t+1 \mid \mathbf{c}_t).
\end{aligned}
\end{equation}
where:
\begin{equation}
    F(k \mid \mathbf{c}_t) = \sum_{j < k} \tilde{P}(j \mid \mathbf{c}_t).
\end{equation}
Through this iterative refinement, the interval $[l, h)$ becomes increasingly \textit{narrow} as more symbols are encoded. After processing all $N$ indices, the final interval $[l_N, h_N)$ uniquely represents the entire sequence, and any real number within this range can serve as the compact binary representation transmitted to the receiver.

At the receiver side, the decoder first loads the shared $n$-gram frequency table, which contains the conditional probability model used during encoding. Using this model, the arithmetic decoder reconstructs the latent index sequence in a sequential manner: starting from the initial interval $[0,1)$, it repeatedly finds the sub-interval whose probability range contains the encoded value and updates the interval accordingly. This process continues until all indices are recovered, producing the exact same discrete code map as the encoder. The recovered indices are then mapped to codewords using the shared codebook to obtain the quantized latent representation $\hat{Z}$, which is finally fed into the semantic decoder to generate the reconstructed data.

\begin{figure}[t]
    \centerline{\includegraphics[scale=0.85]{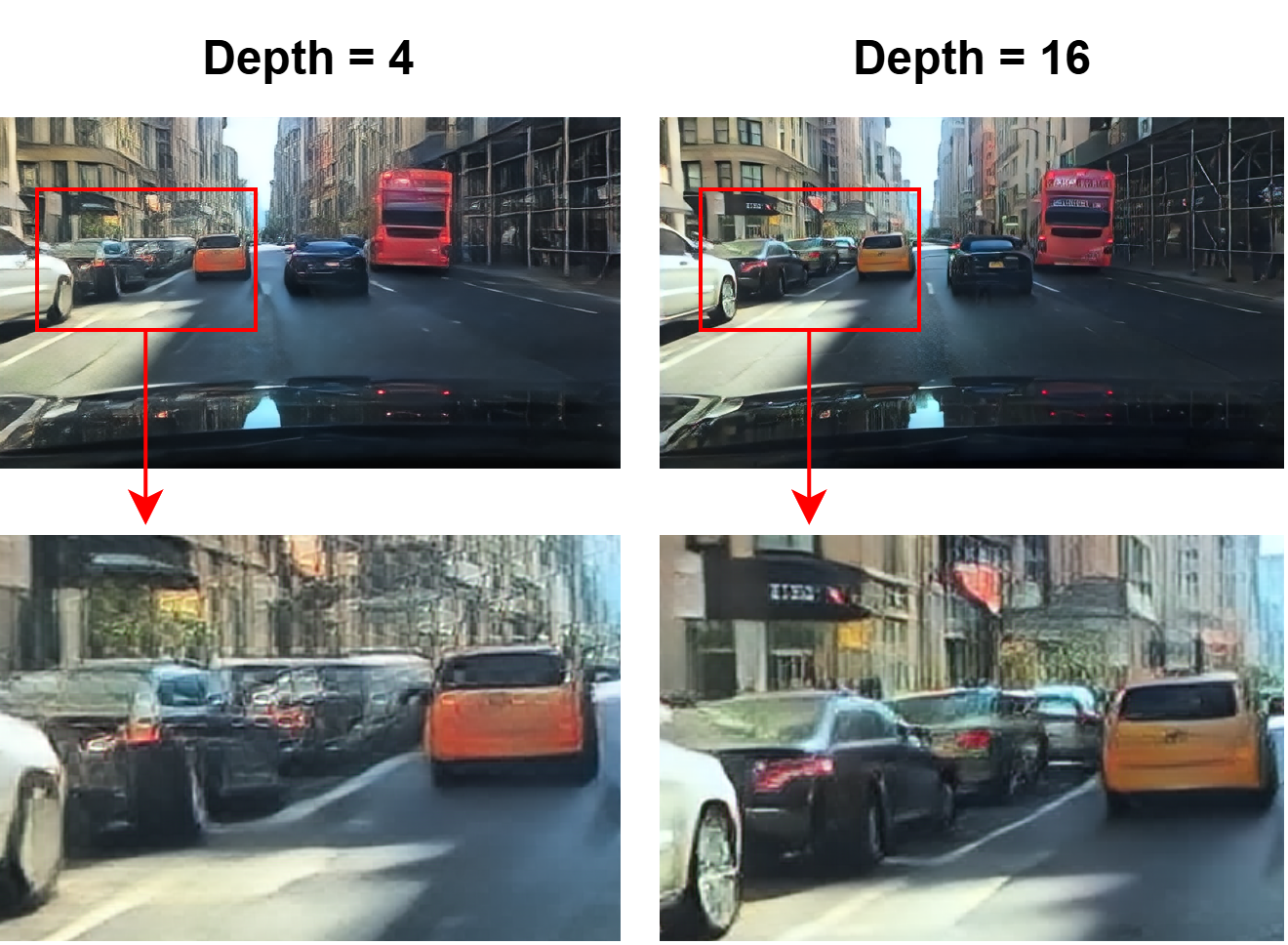}}
    \caption{Visual comparison of reconstructed images under different RQ depths. Increasing the depth preserves richer local details and sharper object contours, especially for nearby vehicles on the road.} 
    \label{fig:qualitative_comparison}
\end{figure}

\section{Adaptive Model Selection under Channel Constraints}
\label{subsec:adaptive_comm}

As illustrated in Fig.~\ref{v2v}, the adaptive depth controller determines the semantic encoding depth according to the observed channel state and system constraints before transmission. This section formalizes this runtime decision process as a channel-aware model selection problem over a discrete set of residual quantization depths.

\subsection{Scaling Dimensions and Trade-offs}

To enable adaptive communication under varying channel bandwidth and latency constraints, the proposed framework exploits two controllable scaling dimensions: the residual quantization depth and the order of the entropy model.

The residual quantization depth serves as the primary runtime adaptation variable. The hierarchical residual quantization structure naturally provides multiple encoding depths, where each additional level progressively refines the residual error from the previous stage. Therefore, deeper quantization preserves finer semantic details and improves reconstruction quality, but produces a larger latent representation and incurs higher transmission and decoding cost. In contrast, shallower quantization provides a more compact representation with lower latency, at the expense of reduced semantic fidelity. As illustrated in Fig.~\ref{fig:qualitative_comparison}, increasing the residual quantization depth leads to clearer reconstructions with richer scene details, sharper vehicle boundaries, and more faithful local structures.

On the other hand, the $n$-gram-based entropy model introduces another trade-off among compression efficiency, computational complexity, and statistical reliability. While increasing $n$ can capture higher-order dependencies among latent indices, the context space grows exponentially, causing data sparsity and rapidly increasing storage and lookup overhead. As shown in our preliminary study of Table~\ref{tab:ngram_overhead}, for RQ-NAC(8), moving from 1-gram to 2-gram only moderately increases encoding latency from 47.2 ms to 77.9 ms, whereas the 3-gram model becomes impractical, requiring 4.4 s encoding latency, 210.7 GiB peak memory, and 22.8 GB storage. Therefore, we treat the entropy model order as an offline tunable parameter and adopt a moderate-order model for real-time deployment. The impact of different $n$-gram orders on compression efficiency is further analyzed in Sec.~\ref{entropy}.

Based on this trade-off analysis, we use the residual quantization depth as a runtime adaptation variable for system configuration, while treating the entropy model order as an offline tunable parameter. This design allows the system to adjust semantic bitrate and reconstruction fidelity according to channel and latency conditions without retraining the semantic codec.

\begin{table}[t]
\centering
\caption{Profiling of different $n$-gram orders for RQ-NAC(8) with a fixed RQ depth of 8.}
\label{tab:ngram_overhead}
\begin{tabular}{c|cccc}
\toprule
$n$ & Enc. (ms) & Dec. (ms) & Peak Mem. & Storage \\
\midrule
1 & 47.2 & 718.5 & 325.3 MiB & 46.6 kB \\
2 & 77.9 & 770.9 & 1.3 GiB & 59.0 MB \\
3 & 4400.8 & 819.2 & 210.7 GiB & 22.8 GB \\
\bottomrule
\end{tabular}
\end{table}

\subsection{Channel-Aware Adaptive Model Selection}

Given the candidate RQ depth set $\mathcal{D}$, we formulate runtime model selection as a discrete optimization problem. At each transmission instance, the encoder observes the effective multicast rate $R_{\mathrm{mc}}$ and the latency budget $T_{\max}$, and selects the semantic configuration that maximizes utility while satisfying the latency constraint.

{Let $d \in \mathcal{D}$ denote a candidate configuration. The adaptive selection problem is formulated as
\begin{equation}
\begin{aligned}
\max_{d \in \mathcal{D}} \quad 
& U(d) = w_q \widetilde{Q}(d) - (1-w_q)\widetilde{T}(d) \\
\text{s.t.} \quad 
& T(d) \leq T_{\max}, \\
& T(d) = T_{\mathrm{enc}}(d) + \mathbb{E}[T_{\mathrm{tx}}(d)] + T_{\mathrm{dec}}(d), \\
& Q(d) = f_Q(d), \\
& d \in \mathcal{D}.
\end{aligned}
\label{eq:adaptive_selection}
\end{equation}
where $Q(d)$ is the offline-profiled semantic reconstruction quality associated with configuration $d$, and $T(d) \triangleq \mathbb{E}[T_{\mathrm{e2e}}(d)]$ denotes the expected end-to-end latency under current channel conditions. The semantic quality $Q(d)$, encoding latency $T_{\mathrm{enc}}(d)$, and decoding latency $T_{\mathrm{dec}}(d)$ are profiled offline for each candidate depth and stored for runtime lookup. Thus, no online quality estimation or ground-truth reference is required during deployment. The profiled \(Q(d)\) represents an average quality operating point obtained from representative vehicular data rather than an image specific estimate.} The normalized terms $\widetilde{Q}(d)$ and $\widetilde{T}(d)$ are obtained from $Q(d)$ and $T(d)$, respectively, so that semantic quality and latency cost can be compared on a consistent scale. The function $f_Q(\cdot)$ represents the offline-profiled mapping from quantization depth to semantic reconstruction quality. The expected transmission latency $\mathbb{E}[T_{\mathrm{tx}}(d)]$ is derived from the block-level HARQ model described in Section~\ref{sec:system_model}, and depends on the compressed bitstream size, the effective multicast rate, and the retransmission process. {The scalar weight $w_q \in [0,1]$ controls the trade-off between semantic fidelity and latency, while the latency weight is given by $1-w_q$. The same weight is applied across all scenarios.}

Based on this formulation, the selection procedure first identifies the feasible candidate set. A candidate depth is considered feasible if its expected end-to-end latency satisfies the deadline constraint:
\begin{equation}
\mathcal{F}=\left\{ d\in\mathcal{D} \;\middle|\; T(d)\le T_{\max}\right\}.
\end{equation}
If no candidate satisfies the above condition, the system falls back to the depth with the minimum expected end-to-end latency:
\begin{equation}
d^\star=\arg\min_{d\in\mathcal{D}} T(d),
\end{equation}
which provides graceful degradation under highly constrained channel conditions.

{When the feasible set is nonempty, the selected depth is obtained by maximizing the utility over $\mathcal{F}$:
\begin{equation}
d^\star=\arg\max_{d\in\mathcal{F}}
\left(
w_q\widetilde{Q}(d)-(1-w_q)\widetilde{T}(d)
\right).
\end{equation}}
After the RQ depth is selected, transmission is performed under the stochastic block-level HARQ process, and the realized latency and delivery success are obtained. Algorithm~\ref{alg:model_selection} summarizes the proposed runtime depth selection procedure. Since the candidate set $\mathcal{D}$ is small, the computational complexity is $\mathcal{O}(|\mathcal{D}|)$, making the approach suitable for real-time deployment without retraining or iterative optimization.

\begin{algorithm}[t]
\caption{Runtime Adaptive Depth Selection for RQ-NAC}
\label{alg:model_selection}
\begin{algorithmic}[1]
\Require {Candidate depth set $\mathcal{D}$; multicast rate $R_{\mathrm{mc}}$; latency budget $T_{\max}$}
\Ensure Selected depth $d^\star$

\State $\mathcal{F} \gets \emptyset$
\ForAll{$d \in \mathcal{D}$}
    \State {retrieve semantic quality $Q(d)$}
    \State estimate expected end-to-end latency
    \[
    \mathbb{E}[T_{\mathrm{e2e}}(d)]
    =
    T_{\mathrm{enc}}(d)+\mathbb{E}[T_{\mathrm{tx}}(d)]+T_{\mathrm{dec}}(d)
    \]
    \If{$\mathbb{E}[T_{\mathrm{e2e}}(d)] \le T_{\max}$}
        \State $\mathcal{F} \gets \mathcal{F} \cup \{d\}$
    \EndIf
\EndFor

\If{$\mathcal{F}=\emptyset$}
    \State $d^\star \gets \arg\min\limits_{d\in\mathcal{D}} \mathbb{E}[T_{\mathrm{e2e}}(d)]$
    \State \Return $d^\star$
\EndIf

\State compute normalized quality score $\widetilde{Q}(d)$ and normalized latency cost $\widetilde{T}(d)$ for all $d\in\mathcal{F}$
\State {$d^\star \gets \arg\max\limits_{d\in\mathcal{F}} \left(
w_q\widetilde{Q}(d)-(1-w_q)\widetilde{T}(d)
\right)$}
\State \Return $d^\star$
\end{algorithmic}
\end{algorithm}

\subsection{Safety-Critical Perception with Domain Adaptation}
\label{subsec:perception_resilience}

While the runtime adaptation relies on a lightweight perceptual quality proxy $Q(d)$ for efficient RQ depth selection, vehicular communication must ultimately be evaluated by its ability to preserve safety-critical semantic information under dynamic wireless conditions. In particular, the perception of nearby vehicles is of primary importance in autonomous driving scenarios, as close-range objects are directly related to collision avoidance and immediate decision-making. Therefore, beyond reconstruction fidelity, it is essential to assess whether the transmitted semantic representations remain effective for downstream perception tasks such as object detection.

A key challenge is the domain gap between generic natural-image data and vehicular driving scenes, as well as the discrepancy between original and reconstructed images. A detector trained on standard datasets may not generalize well to semantic reconstructions that exhibit quantization artifacts and texture smoothing.
To address this issue, we adopt a domain-adaptive fine-tuning strategy based on a YOLOv8n detector as an example. Specifically, the detector is fine-tuned on a vehicle-centric dataset to better capture the structural characteristics of real-world driving scenes. 
{The model is initialized from COCO-pretrained weights and optimized using the standard YOLOv8 detection loss~\cite{yolov8_ultralytics}:
\begin{equation}
\mathcal{L}_{\mathrm{det}}
=
\lambda_{\mathrm{box}}\mathcal{L}_{\mathrm{box}}
+
\lambda_{\mathrm{cls}}\mathcal{L}_{\mathrm{cls}}
+
\lambda_{\mathrm{dfl}}\mathcal{L}_{\mathrm{dfl}},
\end{equation}
where $\mathcal{L}_{\mathrm{box}}$ is based on Complete IoU (CIoU) \cite{zheng2020distance}, $\mathcal{L}_{\mathrm{cls}}$ is implemented using binary cross-entropy with logits, and $\mathcal{L}_{\mathrm{dfl}}$ denotes the Distribution Focal Loss \cite{li2020generalized} used for bounding-box regression.}

Importantly, we do not fine-tune the detector on compressed images. Our approach is motivated by the observation that RQ progressively refines the latent representation and tends to preserve object-level layouts and contours, while fine-grained textures may be smoothed under low-depth compression. This observation is also visually illustrated in Fig.~\ref{fig:qualitative_comparison} and empirically evaluated in Sec.~\ref{subsec:perception_eval}. By fine-tuning the detector on vehicle-centric original images, we encourage it to better capture driving-scene structures and improve its generalization to reconstructed images across different compression levels, without requiring compression-specific retraining.

\begin{table}[t]
\centering
\caption{PHY-layer Parameters.}
\label{tab:phy_params}
\begin{tabular}{l c}
\toprule
Scheduling interval $T_{\mathrm{slot}}$ & 1 ms \\
OFDM symbols per slot & 14 \\
Full subcarriers $N_{\mathrm{sc}}^{\mathrm{full}}$ & 1024 \\
Streams per user & 1 \\
Target BLER & 0.10 \\
Transmission block size & 48 Kbits (6 KB) \\
Maximum retransmissions & 4 \\
MCS table index & 1 \\
\bottomrule
\end{tabular}
\end{table}

\section{Evaluation Methodology}

\subsection{Datasets and Baselines}

{We evaluate RQ-NAC in a bandwidth-limited vehicular communication setting using the 100K Real-World Vehicle Dashcam Image Dataset~\cite{dashcam_dataset}. 
We use a training pool of 20,000 images and 1,000 images for testing. Unless otherwise specified, the RQ-NAC and VQ-VAE models are trained using a 5,000-image subset. We additionally train RQ-NAC(4) using the full 20,000 training set to study the effect of training scale. 
To evaluate generalization under distribution shift, we additionally conduct a case study using 1,000 frames from the Car Crash Dataset~\cite{car_crash_dataset}, which is not used during training.}

{We compare RQ-NAC with different residual quantization depths against VQ-VAE, the recent learning-based compression methods DiffEIC \cite{li2024toward} and QRes-VAE \cite{duan2023lossy}, and the traditional non-learned codecs HEVC and JPEG. Input images, originally sized $1,280{\times}720$, are zero-padded along the height to $1,280{\times}736$ so that the spatial dimensions are compatible with the encoder downsampling operations. The original image dimensions are used for BPP calculation.}

\subsection{Implementation Details}

{The semantic encoder and decoder follow the RQ-VAE architecture as in~\cite{lee2022autoregressive}. The encoder progressively downsamples the RGB input using convolutional and residual blocks and maps it to a 256-dimensional latent representation. The residual quantization module uses a codebook with 2,048 entries and an embedding dimension of 256. Different RQ depths correspond to different numbers of residual refinement stages, providing multiple bitrate and reconstruction fidelity operating points through the hierarchical residual quantization structure. The decoder mirrors the encoder structure and progressively upsamples the quantized latent representation to reconstruct the RGB image.}

{Unless otherwise specified, the NAC module uses a 2-gram context model with smoothing coefficient $\alpha=0.1$. All RQ-NAC models are trained using eight NVIDIA A100 GPUs on the NVIDIA Saturn Cloud platform. 
The n-gram scalability profiling in Table~\ref{tab:ngram_overhead} is conducted on a high-memory Google Cloud instance equipped with an Intel Xeon (Skylake) CPU at approximately 2 GHz and 624 GB RAM, since the 3-gram configuration requires substantially larger memory capacity.
Runtime profiling is conducted on a local computing platform equipped with an Intel Core Ultra 9 275HX CPU and an NVIDIA RTX 5080 Laptop GPU with 16 GB memory. }

\subsection{Evaluation Metrics}
\label{metrics}

\textbf{Reconstruction metrics.} To evaluate the perceptual quality of reconstructed image data, we adopt four data reconstruction metrics in different dimensions. 
The Structural Similarity Index Measure (SSIM) measures the structural and luminance similarity between the reconstructed and ground-truth images. 
The Peak Signal-to-Noise Ratio (PSNR) quantifies pixel-level reconstruction accuracy based on mean squared error (MSE). 
The Learned Perceptual Image Patch Similarity (LPIPS) evaluates perceptual similarity using deep feature representations. 
{Finally, the Fréchet Inception Distance (FID) \cite{heusel2017gans} measures the distributional distance between reconstructed and reference images in the feature space. It is defined as
\begin{equation}
\mathrm{FID}
=
\left\|\mu_r-\mu_g\right\|_2^2
+
\operatorname{Tr}
\left(
\Sigma_r+\Sigma_g
-2(\Sigma_r\Sigma_g)^{1/2}
\right),
\end{equation}
where $\mu_r$ and $\Sigma_r$ denote the mean and covariance of the feature representations of the reference images, while $\mu_g$ and $\Sigma_g$ denote those of the reconstructed images.}

\begin{table}[t]
\centering
\caption{{Scenario Parameters for Multicast Evaluation}}
\label{tab:scenario_params}
\setlength{\tabcolsep}{5pt}
\begin{tabular}{l c c c c c c}
\toprule
\textbf{Scenario} & $T_{\max}$ & $N$ & $\mu_\gamma$ & $\sigma_\gamma$ & $\rho$ & $K$ \\
                  & (ms) & (min--max) & (dB) & (dB) &  & (min--max) \\
\midrule
Cruise    & 1000 & 1--5    & 15 & 4 & 0.3 & 1--3  \\
Junction  & 750  & 10--20  & 10 & 5 & 0.5 & 1--6  \\
Collision & 500  & 50--200 & 8  & 6 & 1.0 & 1--20 \\
\bottomrule
\end{tabular}
\end{table}

\begin{table*}[t]
\centering
\caption{Performance comparison under different compression configurations. 
``$\pm$'' indicates the standard deviation computed over the test set.
\textbf{Bold} and \underline{underlined} values indicate the best and second-best results among \textbf{learned} models, respectively. 
JPEG* is a \textbf{non-learned} codec included as a reference baseline.}
\label{tab:compression_results}
\begin{tabular}{c|cccc|cc|c}
\toprule
\textbf{Configuration} & \textbf{SSIM$\uparrow$} & \textbf{PSNR (dB)$\uparrow$} & \textbf{LPIPS$\downarrow$} & \textbf{FID$\downarrow$} & \textbf{BPP$\downarrow$} & \textbf{CR$\uparrow$} (w/o AC) & \textbf{CR$\uparrow$} (w/ AC)\\

\midrule
VQ-VAE-5k      & 0.7047 $\pm$ 0.1028 & 22.42 $\pm$ 2.89 & 0.2667 $\pm$ 0.0758 & 39.1660 & \textbf{0.01098} & \textbf{2,185.61} & \textbf{2,166.38} \\
\midrule
RQ-NAC(4)-5k     & 0.7400 $\pm$ 0.1059 & 23.57 $\pm$ 2.62 & 0.2263 $\pm$ 0.0692 & 32.8649 & \underline{0.04392} & \underline{546.40} & \underline{671.09}\\
RQ-NAC(8)-5k     & 0.7637 $\pm$ 0.0874 & 24.80 $\pm$ 3.01 & 0.1821 $\pm$ 0.0579 & 26.3956 & 0.08784 & 273.20 & 356.25 \\
RQ-NAC(16)-5k    & \textbf{0.7832} $\pm$ 0.0654 & \textbf{25.00} $\pm$ 2.26 & \textbf{0.1519} $\pm$ 0.0424 & \underline{24.8398} & 0.17569 & 136.60 & 179.32 \\
\midrule
RQ-NAC(4)-20k & \underline{0.7779} $\pm$ 0.0858 & \underline{24.92} $\pm$ 3.29 & \underline{0.1541} $\pm$ 0.0451 & \textbf{22.4336} & 0.04392 & 546.40 & 627.81 \\
\midrule
JPEG* (original dataset)  & — & — &—  & —
&  0.5107 & 47.00 & —\\
\bottomrule

\end{tabular}
\end{table*}

\textbf{Compression metrics.} To evaluate the compression efficiency and data compactness, we employ bits per pixel (BPP) and compression ratio (CR). 
{The BPP represents the average number of bits required to encode each pixel in the compressed image, i.e., 
$\text{BPP} = \frac{b_{\text{comp}}}{H \times W}$,
where $b_{\mathrm{comp}}$ denotes the total number of bits in the compressed bitstream, and $H$ and $W$ denote the original image height and width, respectively.}
The CR is defined as the ratio between the uncompressed and compressed data sizes, i.e., $\text{CR} = \frac{\text{S}_{\text{orig}}}{\text{S}_{\text{comp}}}$,  where $\text{S}_{\text{orig}}$ and $\text{S}_{\text{comp}}$ denote the sizes of the original and compressed data, respectively.
{We also report the native BPP and compression ratio of the JPEG source images provided by the dataset. Since these files are used as the reference inputs rather than reconstructed outputs, reconstruction metrics are not reported for the JPEG baseline.}

\textbf{Communication metrics.}
We evaluate system-level performance using success ratio, deadline-meet ratio, and end-to-end latency. The success ratio is defined as $P_{\mathrm{succ}} = N_{\mathrm{succ}} / N_{\mathrm{tot}}$, where $N_{\mathrm{succ}}$ and $N_{\mathrm{tot}}$ denote the number of successful and total transmissions, respectively. The deadline-meet ratio is defined as $P_{\mathrm{ddl}} = N_{\mathrm{ddl}} / N_{\mathrm{tot}}$, where $N_{\mathrm{ddl}}$ counts transmissions completed within the latency constraint $T_{\max}$. The end-to-end latency comprises the delays incurred by source encoding, wireless transmission, and destination decoding.

\textbf{Perception metrics.}
We evaluate object detection performance using precision, recall, and mean average precision (mAP). To further characterize the relative importance of detected objects in vehicular scenes, we also use a weighted detection score (WDS), defined as
\begin{equation}
\text{WDS} = \frac{\sum_i w_i d_i}{\sum_i w_i},
\end{equation}
where $d_i \in \{0,1\}$ indicates whether the $i$-th ground-truth object is correctly detected, and $w_i$ denotes its importance weight. In this work, $w_i$ is defined as the normalized bounding-box area, i.e., $w_i = \frac{\text{width}_i \times \text{height}_i}{H \times W}$,
{which serves as an importance heuristic by assigning higher weights to the meaningful vehicles occupying larger image regions.} An object is considered correctly detected if it is matched with a prediction of the same class with IoU $\geq 0.5$.

\subsection{PHY Abstraction and Simulation Framework}

We implement the multicast link abstraction using the system-level PHY abstraction module in Sionna SYS \cite{hoydis2022sionna}. The module maps the effective SINR and allocated resource fraction to the number of successfully decoded bits per scheduling interval, denoted by $b_{\mathrm{slot}}$. An inner-loop link adaptation mechanism is adopted to select the modulation and coding scheme (MCS) index such that a target BLER is satisfied.
The transmission process follows the block-based model described in Section~\ref{sec:system_model}, where the payload is segmented into fixed-size transmission blocks and delivered using HARQ with a bounded number of retransmissions. The corresponding PHY abstraction parameters are summarized in Table~\ref{tab:phy_params}.

{For each Monte Carlo iteration, we instantiate a multicast scenario by sampling:
(i) the multicast group size $N$ uniformly from a scenario-specific range $[N_{\min},N_{\max}]$,
(ii) per-user SINR values ${\gamma_u}$ from a Gaussian distribution $\mathcal{N}(\mu_{\gamma},\sigma_{\gamma}^2)$ in dB, and
(iii) the number of concurrent multicast flows $K$ uniformly from $[K_{\min},K_{\max}]$.
The effective SINR $\gamma_\eta$ is computed using the coverage-based abstraction defined in Section~\ref{sec:system_model}, and the effective resource fraction is given by $\alpha_{\mathrm{eff}}=\rho/K$. The end-to-end deadline $T_{\max}$ is scenario-dependent, as summarized in Table~\ref{tab:scenario_params}.
In our implementation, the quality and latency terms are normalized within the feasible candidate set using standard min--max normalization. A single quality weight $w_q=0.8$ is used across all scenarios, with the corresponding latency weight given by $1-w_q=0.2$.}

{Note that the adopted channel model is intended as a system-level abstraction rather than a time-resolved mobility-aware channel model. Accordingly, temporally correlated fast fading, mobility-induced Doppler, detailed inter-cell or co-channel interference, and CSI estimation or feedback delay are not explicitly modeled, and the controller assumes that the current effective SINR estimate is available at the time of depth selection.}

\begin{figure}[t]
    \centerline{\includegraphics[scale=0.38]{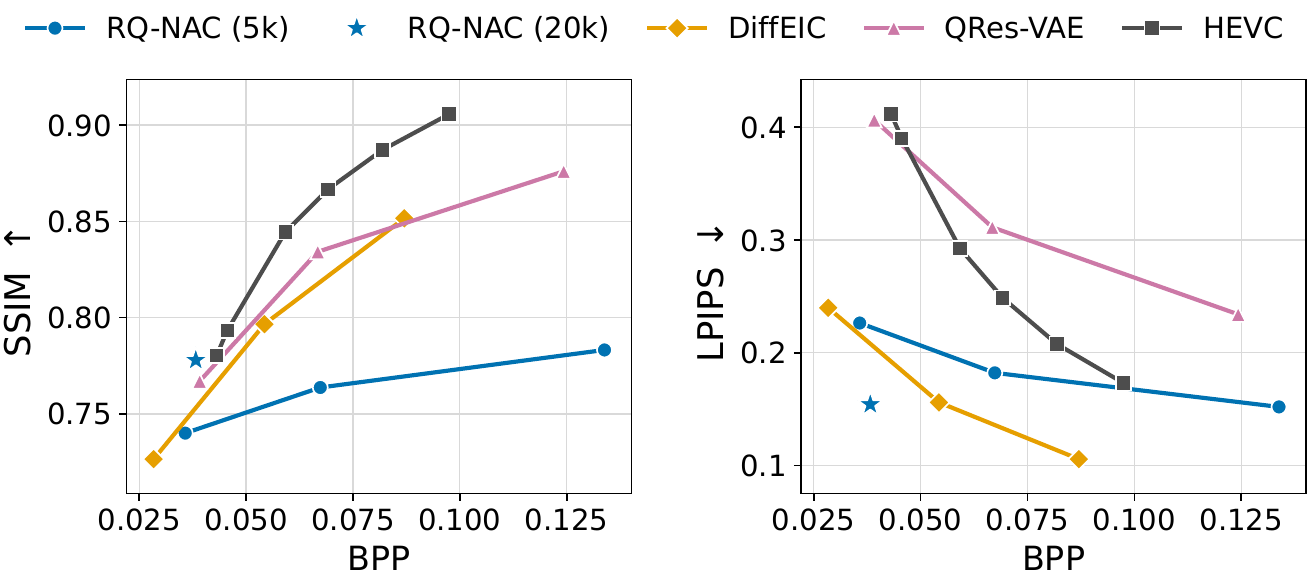}}
\caption{Rate-distortion comparison of RQ-NAC and representative image compression baselines on the same test set.}
    \label{fig:rd_comparison}
\end{figure}

\begin{figure*}[t]
    \centerline{\includegraphics[scale=0.88]{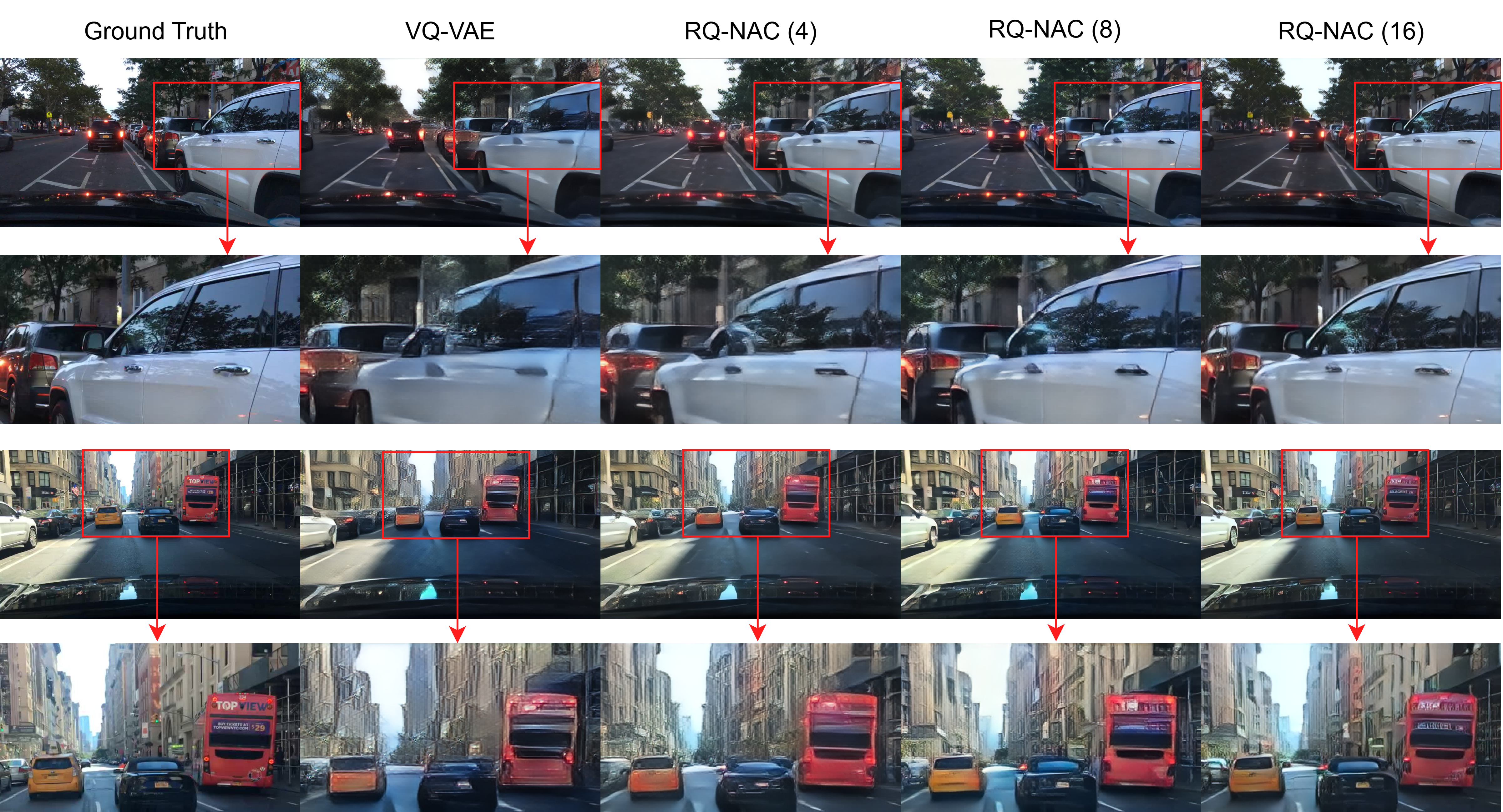}}
    \caption{Qualitative comparison of reconstructed images produced by different compression models. 
    The proposed RQ-NAC progressively improves visual fidelity as the residual quantization depth increases, recovering sharper edges, clearer object contours, and more stable scene structures. This qualitative trend is consistent with the quantitative results in Table IV.} 
    \label{iov}
\end{figure*}

\section{Experimental Results}
This section evaluates the proposed RQ-NAC framework under bandwidth-limited vehicular communication scenarios. We first examine its semantic reconstruction quality, compression efficiency, and entropy coding performance under different residual quantization depths. We then evaluate the runtime depth adaptation mechanism under dynamic channel conditions, followed by downstream object detection and out-of-distribution accident-scene case studies to verify perception utility and robustness.

\subsection{Performance of Semantic Compression}
\label{subsec:perception_eval}

\subsubsection{Performance of Residual Quantization}
First, we compare the proposed RQ-NAC with traditional codecs and learning-based baselines, as shown in Table~\ref{tab:compression_results}. We include VQ-VAE as a learning-based baseline, which can be viewed as a single-stage quantization model.
Overall, we observe a clear trend that increasing the quantization depth consistently improves reconstruction quality across all perceptual and distortion metrics. 
For example, RQ-NAC(16) achieves the best performance among learned models trained on the 5k subset, reaching an SSIM of 0.7832, PSNR of 25.00 dB, and LPIPS of 0.1519, significantly outperforming the VQ-VAE baseline (SSIM 0.7047, PSNR 22.42 dB). 
This improvement comes at the cost of reduced compression efficiency, where the compression ratio decreases from 546.40× at depth 4 to 136.60× at depth 16, reflecting the trade-off between reconstruction quality and bitrate.

To further investigate the impact of training data size, we evaluate a model with quantization depth 4 trained on an expanded dataset of 20k samples. 
As shown in Table~\ref{tab:compression_results}, the RQ-NAC(4)-20k model achieves comparable reconstruction performance to the much deeper RQ-NAC(16)-5k variant, indicating that increasing the amount of training data can enhance the model’s representational capacity and improve the reconstruction quality with lower bitrate overhead.  
Unless otherwise specified, all other models are trained on the 5k subset.

{We further compare RQ-NAC with two recent learning-based image compression methods, DiffEIC~\cite{li2024toward} and QRes-VAE~\cite{duan2023lossy}, as well as HEVC over multiple bitrate operating points, using the same test set. HEVC is evaluated using constant rate factor (CRF) values from 36 to 51. As shown in Fig.~\ref{fig:rd_comparison}, RQ-NAC provides competitive rate-distortion performance in the ultra-low-bitrate regime. In particular, after increasing the training set from 5k to 20k samples, RQ-NAC(4) achieves an SSIM of 0.7779 and an LPIPS of 0.1541 at approximately 0.04 BPP. At this bitrate, RQ-NAC(4)-20k achieves substantially lower LPIPS than the considered learned compression baselines and HEVC, while maintaining competitive SSIM. HEVC requires a considerably higher bitrate to approach a similar perceptual-quality level. This improvement is achieved without changing the quantization depth or representation size, suggesting that the performance of the current RQ-NAC models is strongly influenced by the available training scale rather than being fundamentally limited by the residual-quantization architecture.}

Fig.~\ref{iov} presents a qualitative comparison of reconstructed images produced by different models. The VQ-VAE baseline exhibits noticeable distortion and global blurriness, which degrade perceptual quality and hinder semantic interpretation of scene details such as vehicle contours and road boundaries. In contrast, the proposed RQ-NAC with depth 4 effectively preserves the overall scene structure and eliminates most visible artifacts, although the reconstruction remains slightly blurred due to limited quantization depth. Increasing the residual quantization depth further enhances visual fidelity: RQ-NAC(8) and RQ-NAC(16) recover finer textures and sharper edges, producing images that are perceptually closer to the ground truth. These results confirm that hierarchical residual quantization enables a clear trade-off between compression efficiency and perceptual quality, with deeper models achieving more faithful semantic reconstruction.

\begin{figure}[t]
  \centering
  \subfloat[Impact of $n$-gram order. \label{fig:ngram}]{
    \includegraphics[width=0.46\linewidth]{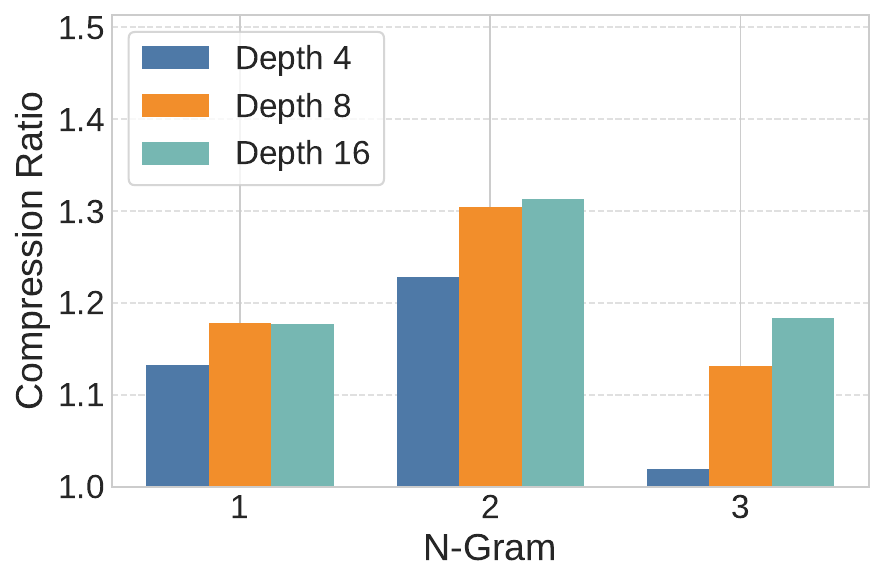}
  }
  \hfill
  \subfloat[Impact of coefficient $\alpha$. \label{fig:alpha}]{
    \includegraphics[width=0.46\linewidth]{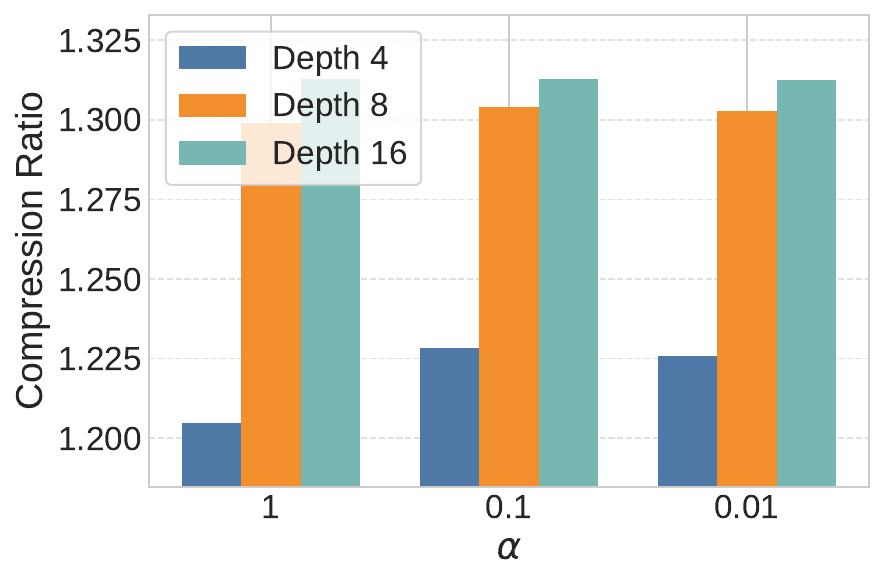}
  }
  \caption{Parametric study on compression ratio under different quantization depths.}
  \label{fig:compression-ratio}
  \vspace{-1em}
\end{figure}

\subsubsection{Performance of Entropy-based Lossless Compression}
\label{entropy}
We then evaluate the effectiveness of the proposed context-adaptive entropy coding model under different $n$-gram orders and quantization depths. 
Fig.~\ref{fig:ngram} presents the resulting stage-2 compression ratios across different quantization depths.
We observe that even without the contextual modeling ($n=1$), applying AC directly yields a notable compression ratio of approximately $1.15\times$, demonstrating the redundancy present in the latent space. 
When incorporating a 2-gram context model, the compression ratio improves significantly to around $1.3\times$, indicating that short-range dependencies between adjacent code symbols can be further leveraged to enhance compression performance.
However, the 3-gram case leads to a degradation in compression efficiency because the estimated empirical frequencies $\tilde{P}(s_t \mid \mathbf{c}_t)$ deviate from the true probabilities $P(s_t \mid \mathbf{c}_t)$ under \textit{limited data}. 
The exponentially growing context space causes such data sparsity, leading to unreliable and nearly uniform estimates that increase entropy and reduce compression efficiency.
In theory, higher-order $n$-grams can yield better compression as they more accurately model local dependencies, but require significantly larger datasets to avoid context sparsity and maintain reliable probability estimates. 
Besides, both the storage and lookup cost for frequency tables grow exponentially with $n$, which can quickly become impractical for resource-constrained applications.

To mitigate sparsity, we evaluate additive smoothing with $\alpha \in \{1, 0.1, 0.01\}$ under the 2-gram model. As shown in Fig.~\ref{fig:alpha}, varying $\alpha$ results in only marginal compression differences across quantization depths, with $\alpha = 0.1$ slightly outperforming others at depth of 4 and 8. We therefore adopt $\alpha = 0.1$ as the default, and all “w/ AC” results in Table~\ref{tab:compression_results} use this configuration. While VQ-VAE may exhibit degraded performance with n-gram modeling due to its relatively short latent sequences, which limit the effectiveness of context-based estimation and can introduce estimation bias, 2-gram modeling is most effective for our RQ-NAC,  achieving up to \textbf{671.09×} compression at depth of 4 with high perceptual quality.

\subsection{Adaptive Performance under Dynamic Channel Conditions}

\begin{table}[t]
\centering
\caption{Profiling results of compression and restoration.}
\label{tab:profiling}
\begin{tabular}{c|cc|cc}
\toprule
\multirow{2}{*}{\textbf{Depth}} & \multicolumn{2}{c|}{\textbf{Compress (ms)}} & \multicolumn{2}{c}{\textbf{Restore (ms)}} \\
\cmidrule(lr){2-3} \cmidrule(lr){4-5}
 & \textbf{RQ-VAE} & \textbf{NAC} & \textbf{RQ-VAE} & \textbf{NAC} \\
\midrule
RQ-NAC(4)  & 119.0571 & 15.7462 & 196.3354 & 114.7885 \\
RQ-NAC(8)  & 120.4951 & 22.8375 & 195.2968 & 205.2417 \\
RQ-NAC(16) & 120.2094 & 36.4351 & 195.3817 & 475.0927 \\
\bottomrule
\end{tabular}
\vspace{-0.5em}
\end{table}

\begin{figure}[t]
  \centering
  \captionsetup[subfloat]{width=0.33\linewidth}

  \subfloat[{Success and deadline-meet ratios.}\label{fig:deadline}]{%
    \includegraphics[width=0.33\linewidth]{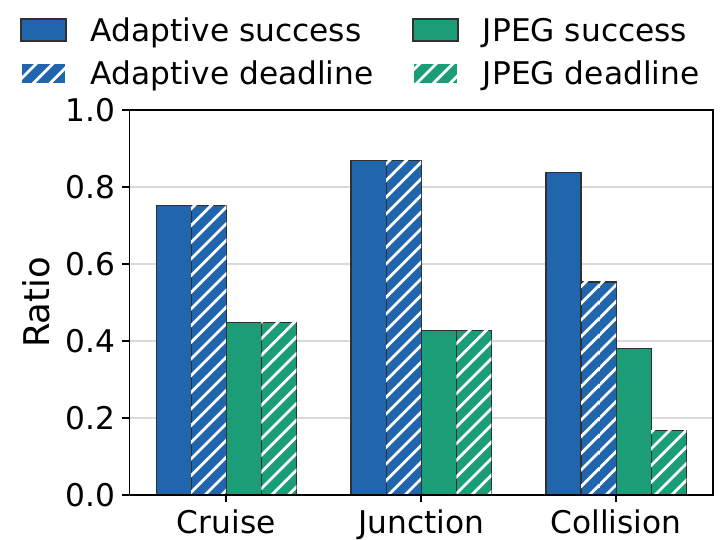}%
  }\hfill%
  \subfloat[{Average end-to-end latency.}\label{fig:avg_e2e}]{%
    \includegraphics[width=0.33\linewidth]{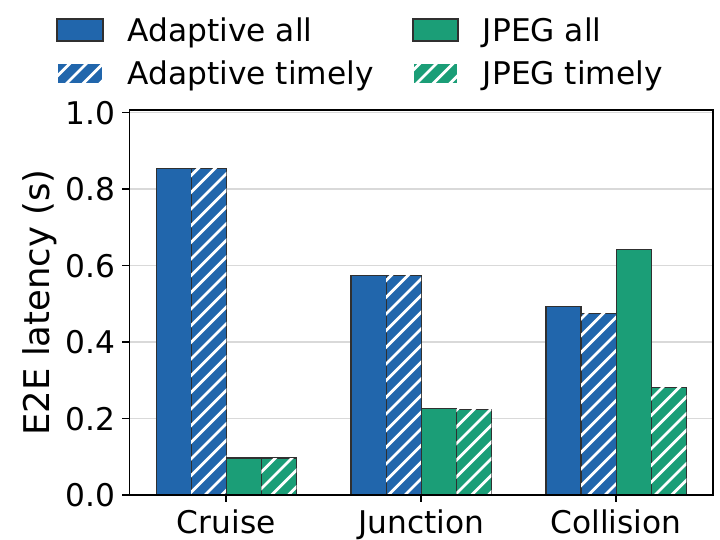}%
  }\hfill%
  \subfloat[{Average transmission latency.}\label{fig:goodput}]{%
    \includegraphics[width=0.33\linewidth]{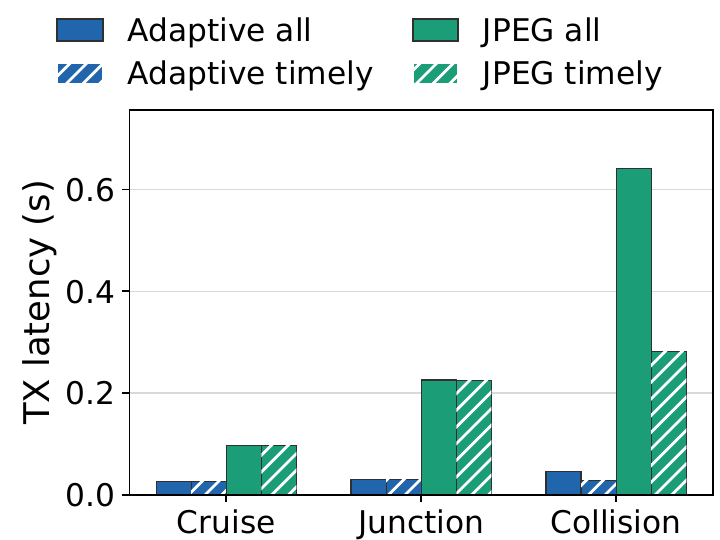}%
  }

  \caption{{Performance comparison between the proposed model adaptation method and the baseline under the three scenarios. “success” indicates the completion success ratio, and “deadline” indicates the deadline-meet ratio. “all” refers to averages computed over all successful transmissions, whereas “timely” refers to averages computed over deadline-satisfied transmissions. }}
  \label{fig:model_adaptation}
\end{figure}

\subsubsection{Profiling Results and System Efficiency}

{To evaluate the practical deployment cost of the proposed SemCom pipeline, we first profile the runtime performance of both the compression and restoration modules under an edge computing setting. 
The dataset provides source images in JPEG format, which serve as the common source data for both methods. The JPEG baseline directly transmits the provided JPEG files, while RQ-NAC decodes them into RGB images before semantic encoding. 
The reported runtime for RQ-NAC includes the computational overhead of semantic compression and decoding modules. }

\begin{table}[t]
\centering
\caption{{Model selection ratio under different vehicular scenarios.}}
\label{tab:model_selection}
\begin{tabular}{c|ccc}
\toprule
\multirow{2}{*}{\textbf{Scenario}} & \multicolumn{3}{c}{\textbf{Selected Model}} \\
\cmidrule(lr){2-4}
 & \textbf{RQ-NAC(4)} & \textbf{RQ-NAC(8)} & \textbf{RQ-NAC(16)} \\
\midrule
Cruise    & 0.4\%   & 0.0\%  & 99.6\% \\
Junction  & 3.1\%   & 96.9\% & 0.0\%  \\
Collision & 100.0\% & 0.0\%  & 0.0\%  \\
\bottomrule
\end{tabular}
\end{table}

Table~\ref{tab:profiling} reports the average per-image processing latency of each component, including the RQ-VAE-based semantic encoder/decoder and the NAC entropy coding/decoding modules, across different RQ depths. We observe that the semantic encoding time remains relatively stable (approximately 120 ms) across different RQ depths. This is because the main computational cost of semantic encoding lies in the convolutional backbone used to extract the latent feature map, which is executed only once per image and does not scale with the RQ depth. A larger RQ depth mainly introduces additional codeword lookup and residual index generation on the latent features, whose overhead is relatively minor compared with the backbone inference.

In contrast, the entropy coding overhead increases with the RQ depth due to the longer latent index sequences, particularly on the decoder side where the NAC part restoration becomes the dominant cost at higher RQ depths.
Overall, the profiling results show that the RQ-VAE encoding latency remains nearly constant across RQ depths, whereas the NAC decoding cost increases with the length of the latent index sequence. Even at higher depths, the system maintains tractable latency at hundreds of millisecond level, making it suitable for near-real-time deployment in bandwidth-constrained vehicular network environments. 

\begin{figure}[t]
    \centerline{\includegraphics[scale=0.45]{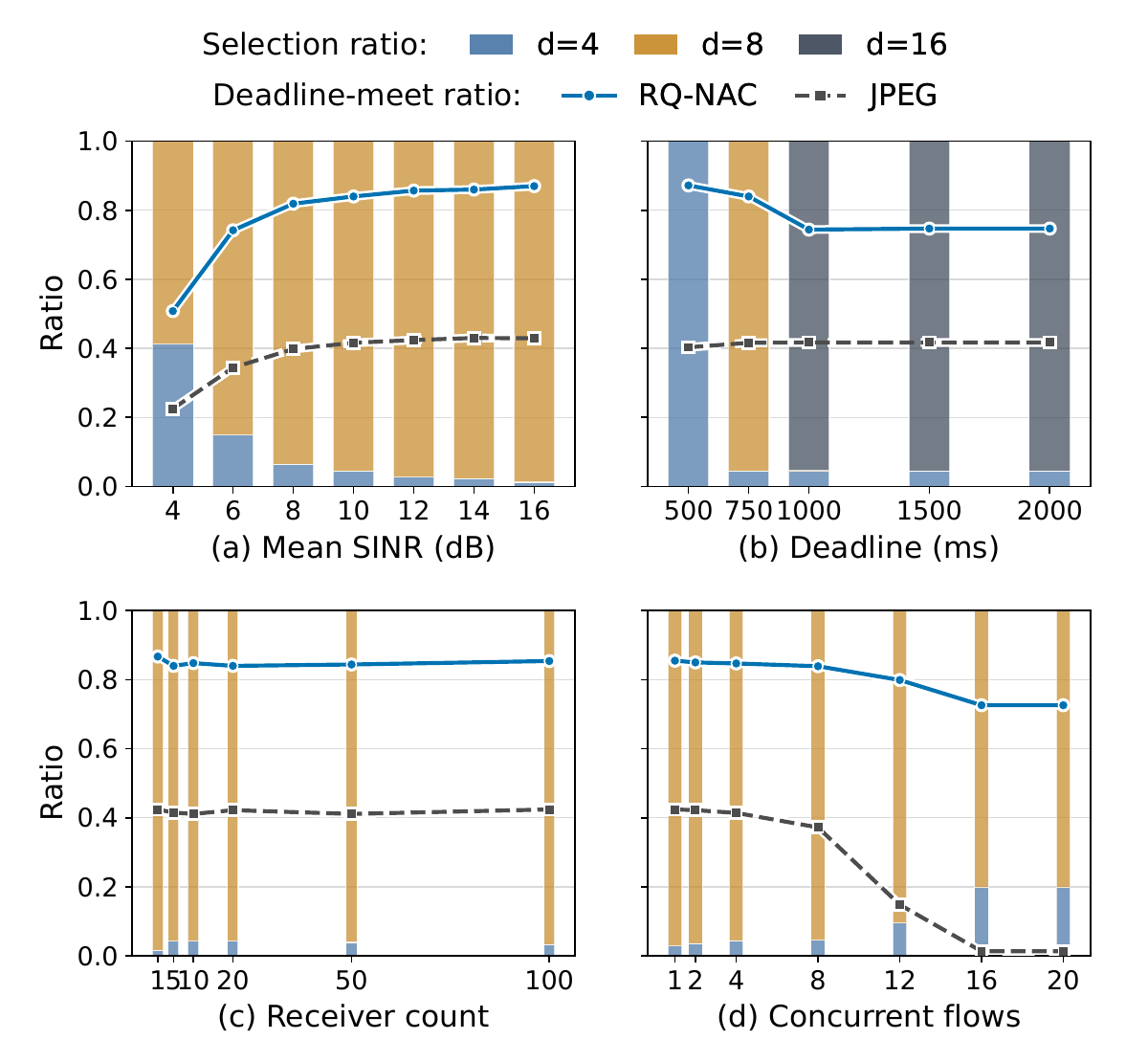}}
    \vspace{-1em}
    \caption{Controller behavior under controlled variations of channel and system conditions in the \textbf{Junction} scenario.}
    \label{fig:controller_sweeps}
\vspace{-0.5em}
\end{figure}

\begin{table*}[t]
\centering
\caption{Vehicular object detection performance under different reconstruction depths using pretrained and fine-tuned YOLOv8n detectors on the Test110 dataset. In addition to standard metrics, we report a Weighted Detection Score (WDS), where object importance is approximated by bounding-box area.}
\label{tab:yolo_detection_full}
\begin{tabular}{c|cccc|ccccc}
\toprule
\multirow{2}{*}{\textbf{Configuration}} 
& \multicolumn{4}{c|}{\textbf{Pretrained YOLOv8n}} 
& \multicolumn{5}{c}{\textbf{Fine-tuned YOLOv8n}} \\
\cline{2-10}
& \rule{0pt}{2.5ex}\textbf{Prec.} & \textbf{Rec.} & \textbf{mAP50} & \textbf{mAP50-95}
& \textbf{Prec.} & \textbf{Rec.} & \textbf{mAP50} & \textbf{mAP50-95} & \textbf{WDS} \\
\midrule
VQ-VAE       
& 0.695 & 0.467 & 0.516 & 0.298 
& 0.778 & 0.615 & 0.673 & 0.398 & 0.767 \\
\midrule
RQ-NAC(4)       
& 0.631 & 0.631 & 0.621 & 0.394 
& 0.765 & 0.714 & 0.785 & 0.502 & 0.820 \\

RQ-NAC(8)       
& 0.675 & 0.674 & 0.689 & 0.475 
& 0.819 & 0.747 & 0.843 & 0.578 & 0.886 \\

RQ-NAC(16)      
& \underline{0.710} & \underline{0.739} & \underline{0.733} & \underline{0.523}
& \textbf{0.861} & \underline{0.781} & \underline{0.880} & \underline{0.633} & \underline{0.897} \\

\midrule
JPEG (Original)
& \textbf{0.752} & \textbf{0.757} & \textbf{0.761} & \textbf{0.575} 
& \underline{0.832} & \textbf{0.868} & \textbf{0.929} & \textbf{0.691} & \textbf{0.932} \\
\bottomrule
\end{tabular}
\end{table*}

\subsubsection{Performance of Model Adaptation}

Fig. \ref{fig:model_adaptation} and Table ~\ref{tab:model_selection} show that the selected RQ depth varies across the three operating scenarios as communication load and latency constraints change. In low-density Cruise scenarios, the vehicle controller predominantly selects the deepest configuration, indicating that the communication condition is sufficient to support higher-fidelity semantic transmission. In Junction scenarios, the selected RQ depth shifts to an intermediate operating point, reflecting the need to balance reconstruction quality and communication efficiency under moderate multicast load. In Collision scenarios, the controller consistently favors the shallowest configuration, prioritizing compactness and fast delivery under the most stringent communication conditions. This trend confirms that the proposed adaptation mechanism in Sec. V is able to adjust the semantic bitrate according to the effective multicast capacity and application deadline, without requiring model retraining or online optimization.

\subsubsection{{Sensitivity Analysis}}
{To further verify that this behavior is not a fixed mapping from scenario labels to quantization depths, Fig.~\ref{fig:controller_sweeps} evaluates the controller within the same Junction scenario while independently varying channel and system conditions. As the available latency budget increases in Fig.~\ref{fig:controller_sweeps}(b), the selected operating point progressively shifts from shallow to deeper quantization, reflecting the increased opportunity to preserve semantic fidelity. A similar tendency is observed as the effective channel quality improves or the number of concurrent flows decreases, as shown in Fig.~\ref{fig:controller_sweeps}(a) and in Fig.~\ref{fig:controller_sweeps}(d), respectively. In contrast, variations in the receiver population have a comparatively smaller effect under the adopted coverage-quantile multicast abstraction (Fig.~\ref{fig:controller_sweeps}(c)), because the receiver population primarily influences the sampled lower-tail SINR rather than directly reducing the per-flow resource fraction. These controlled experiments demonstrate that the selected RQ depth is driven by instantaneous communication constraints rather than by the specific scenario.}

{The latency results further highlight the trade-off between communication efficiency and semantic processing overhead. RQ-NAC substantially reduces transmission latency because its semantic representations require fewer transmitted bits than the baselines. Its end-to-end latency additionally includes neural encoding, decoding, and entropy-coding overhead, which can make the conditional latency of successfully delivered JPEG samples smaller in the less constrained scenarios. However, those JPEG results are obtained from a much smaller subset of transmissions that complete successfully and within the deadline. Under the most stringent Collision setting, RQ-NAC simultaneously provides higher timely-delivery reliability and lower end-to-end latency, illustrating the benefit of adaptive semantic compression when communication resources are severely limited.}

\begin{figure}[t]
    \centerline{\includegraphics[scale=0.36]{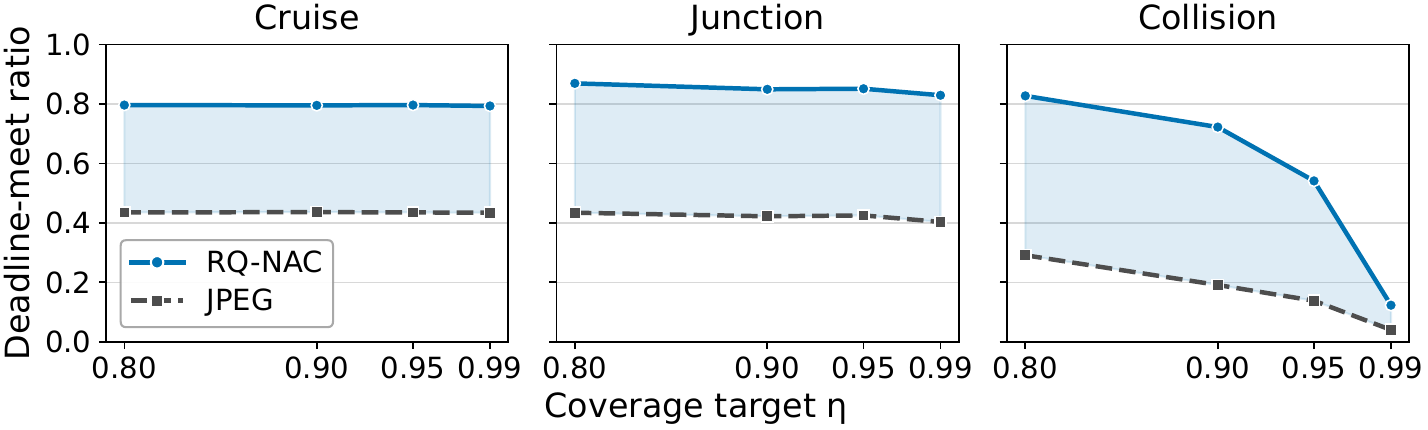}}
    \vspace{-1em}
    \caption{Sensitivity of the deadline-meet ratio to the target receiver coverage ratio $\eta$ under different scenarios.}
    \label{eta}
\vspace{-0.5em}
\end{figure}

{We also investigated the impact of the target receiver coverage ratio $\eta$ on the deadline-meet performance in Fig.~\ref{eta}. As $\eta$ increases, it is observed that the effective multicast SINR becomes more conservative, leading to lower deadline-meet ratios, particularly under the stringent Collision scenario. However, the proposed adaptive RQ-NAC consistently outperforms JPEG across all tested coverage targets, indicating that its relative reliability advantage is preserved under different multicast coverage requirements.}

\subsection{Perception Performance with Semantic Applications}

To evaluate whether the reconstructed semantic images remain useful for downstream perception, we conduct a task-oriented evaluation using a YOLOv8n object detector. We consider two settings: (i) a pretrained detector without adaptation, and (ii) a fine-tuned detector trained on original images captured by vehicles to mitigate domain shift.
We construct a controlled test set consisting of 110 dashcam images, with 1--3 near-field vehicles annotated per image. The same set of images is used across all compression levels to ensure a fair comparison. For the pretrained setting, we directly evaluate a YOLOv8n model pretrained on the COCO dataset~\cite{lin2014microsoft}. For the fine-tuned setting, we train a single-class vehicle detector on a separate dataset of original images with no overlap with the test set, and select the best checkpoint using an independent validation set of 40 images based on validation mAP$_{50-95}$.
Detection performance is evaluated using the metrics defined in Section~\ref{metrics}, including precision, recall, mAP, and WDS.

As shown in Table~\ref{tab:yolo_detection_full}, detection performance improves consistently as the residual quantization depth increases, indicating that deeper RQ-NAC reconstructions preserve more task-relevant visual structures. Under the pretrained detector, RQ-NAC substantially outperforms VQ-VAE and approaches the original JPEG baseline, demonstrating the benefit of hierarchical residual quantization for downstream perception.
After fine-tuning on domain-specific vehicular data, detection performance further improves across all compression levels. In particular, RQ-NAC(16) achieves an mAP$_{50}$ of 0.880 and even surpasses the original baseline in precision, suggesting that the reconstructed semantic images can retain highly discriminative object structures for vehicle detection. RQ-NAC(8) also maintains strong performance, showing that moderate reconstruction depth can provide a favorable balance between compression and perception quality.
{However, the original JPEG baseline retains the highest overall mAP$_{50}$ and mAP$_{50-95}$, indicating a tradeoff between perception fidelity and transmission payload.}

{The WDS results further confirm that higher reconstruction depth better preserves objects assigned higher importance scores. RQ-NAC improves WDS monotonically with depth and reaches 0.897 at depth 16, remaining close to the original baseline. This indicates that the proposed semantic compression retains information associated with vehicles occupying larger image regions when sufficient residual refinement is used.}
These results show that RQ-NAC preserves downstream perception utility under semantic compression, while domain-adaptive fine-tuning further improves robustness to reconstruction artifacts in vehicular perception applications.

\begin{figure}[t]
    \centerline{\includegraphics[scale=0.18]{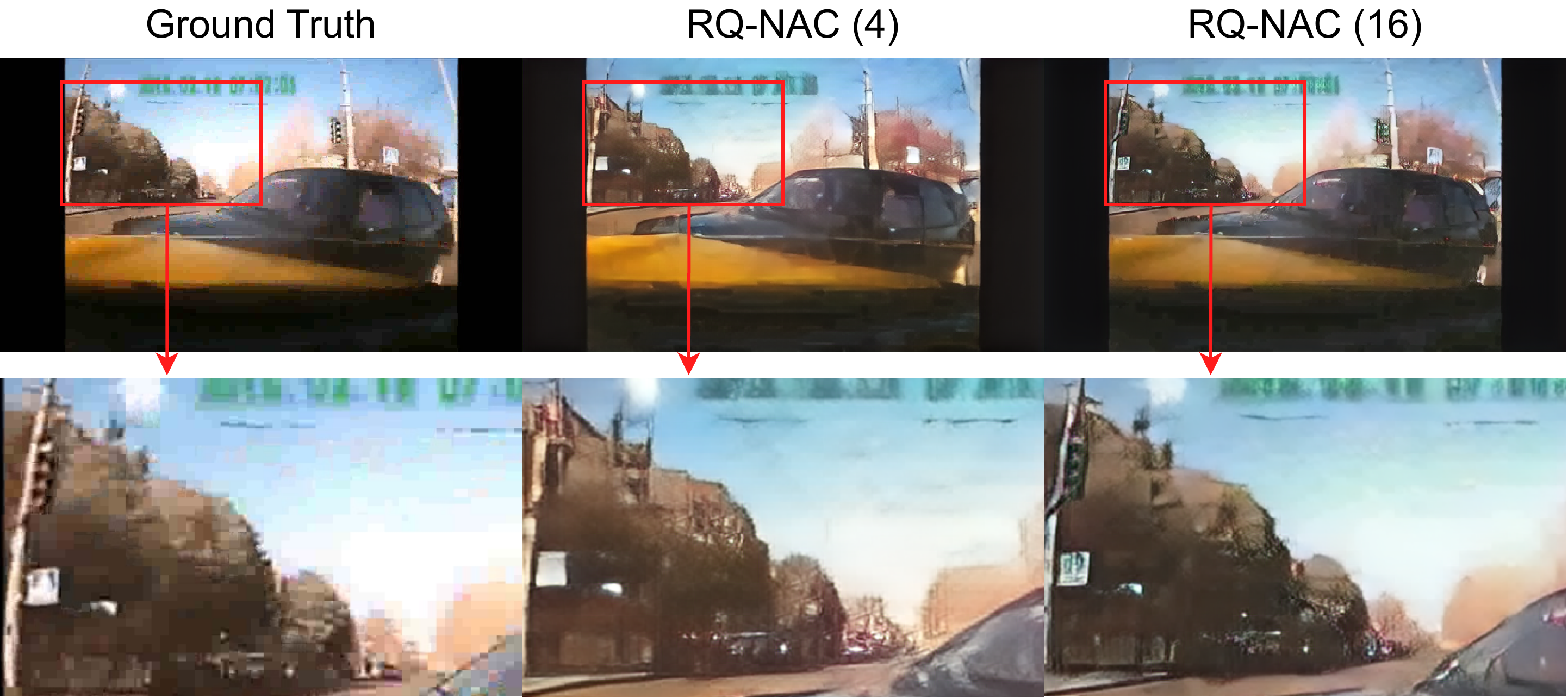}}
    \caption{Case study on unseen accident scenes. RQ-NAC maintains strong reconstruction quality under distribution shift, with deeper quantization preserving finer visual details, including clearer traffic lights and object boundaries.}
    \label{case}
\end{figure}

\subsection{Robustness under Distribution Shift}

{To evaluate the generalization ability of RQ-NAC under distribution shift, we conduct a case study using a new Car Crash Dataset~\cite{car_crash_dataset}, which contains real-world accident scenes not seen during training. We directly apply two RQ-NAC models pretrained on the original vehicular dataset, with RQ depths of 4 and 16, without any fine-tuning.
As shown in Fig.~\ref{case}, both models preserve the main scene structures under this unseen domain. The depth-4 model retains major elements such as vehicle outlines, traffic lights, and road structures, while the depth-16 model reconstructs finer details including object textures and sharper edges.}

{Table~\ref{tab:case} further reports the reconstruction quality together with the additional compression gain achieved by NAC. Since the latent representation dimensions remain unchanged for a given RQ depth, the compression ratio before arithmetic coding is identical to that reported in Table~\ref{tab:compression_results}. We therefore evaluate the robustness of the entropy model by directly applying the same 2-gram frequency table learned from the original training data to the unseen Car Crash Dataset, without updating or re-estimating the table.
The NAC gain remains above 1.2$\times$ for both configurations and is close to the corresponding results on the original test set in Fig. 6, showing that the fixed 2-gram frequency table remains effective on the tested unseen accident scenes without re-estimation.}

\begin{table}[t]
\centering
\caption{Quantitative results of the out-of-distribution case study.}
\label{tab:case}
\begin{tabular}{c|cc|c}
\toprule
\textbf{Configuration} & \textbf{SSIM$\uparrow$} & \textbf{LPIPS$\downarrow$} & \textbf{2-gram AC Gain$\uparrow$} \\
\midrule
RQ-NAC(4)  & 0.6764 & 0.2752 & 1.2137$\times$ \\
RQ-NAC(16) & 0.7369 & 0.2067 & 1.3068$\times$ \\
\bottomrule
\end{tabular}
\end{table}

\section{Conclusion}

In this paper, we proposed a resilient and adaptive visual semantic communication framework for vehicular networks. By integrating residual quantization with a lightweight depth adaptation mechanism, the proposed approach enables flexible trade-offs between semantic fidelity and communication efficiency under dynamic network conditions. In addition, we adopt a perception-aware evaluation framework that emphasizes downstream task utility in safety-critical scenarios.
Through system-level evaluation, the proposed method demonstrates significant improvements in delivery reliability and timely transmission compared with traditional image transmission approaches, while maintaining low-latency operation and robust object detection performance. The results also indicate that adaptive semantic compression can effectively mitigate the impact of channel variability and network congestion, leading to more robust visual information delivery in diverse vehicular application environments.

\section{Acknowledgment}

This research was supported by the National Science Foundation through Award CNS--2440756, CNS--2312138, and NVIDIA Academic Program Grant.

\bibliographystyle{IEEEbib}
\bibliography{references}

\end{document}